\documentclass[manuscript]{aastex}

\usepackage{amsmath}
\usepackage{amssymb}
\usepackage{natbib}
\usepackage{fancyhdr}
\usepackage{graphicx}
\usepackage{lscape}
\usepackage{rotating}
\usepackage{subfigure}
\usepackage{wrapfig}
\usepackage{afterpage}
\usepackage{pifont}
\usepackage{rotating}
\usepackage{subfigure}

\newcommand{\moo}{\rm $\mu$m}

\shorttitle{The origin of the MFIR emission in radio galaxies}
\shortauthors{Dicken et al.}

\begin{document}

\title{The origin of the infrared emission in radio galaxies. III. Analysis of 3CRR objects}

\author{D. Dicken\altaffilmark{1}, C. Tadhunter\altaffilmark{2}, D. Axon\altaffilmark{1,3}, A. Robinson\altaffilmark{1}, R.
Morganti\altaffilmark{4,5}, P. Kharb\altaffilmark{1} }

    \altaffiltext{1}{Department of Physics and Astronomy, Rochester
    Institute of Technology, 84 Lomb Memorial Drive, Rochester NY
    14623, USA ; dxdsps@rit.edu, djasps@rit.edu, axrsps@rit.edu, pxksps@rit.edu}
    \altaffiltext{2}{Department of Physics and Astronomy, University of
    Sheffield, Hounsfield Road, Sheffield, S3 7RH, UK; c.tadhunter@sheffield.ac.uk} 
     \altaffiltext{3}{Department of Physics and Astronomy, University of
    Sussex, Pevensey 2, University of Sussex, Falmer, Brighton, BN1 9QH, UK} 
    \altaffiltext{4}{ASTRON, P.O. Box 2,
    7990 AA Dwingeloo The Netherlands; morganti@astron.nl}
    \altaffiltext{5}{Kapetyn Astronmical Institute, University of Groningen, Postbuss 800, 9700 AV Groningen, The Netherlands}

\begin{abstract}
We present Spitzer photometric data for a complete sample of 19 low redshift ($z<$~0.1) 3CRR radio galaxies as part  of our efforts to understand the origin of the prodigious mid- to far-infrared (MFIR) emission from radio-loud AGN. Our results show a correlation between AGN power (indicated by [OIII]$\lambda5007$ emission line luminosity) and 24\moo\ luminosity. This result is consistent with the 24\moo\ thermal emission originating from warm dust heated directly by AGN illumination. Applying the same correlation test for 70\moo\ luminosity against [OIII] luminosity we find this relation to suffer from increased scatter compared to that of 24\moo. In line with our results for the higher-radio-frequency-selected 2Jy sample, we are able to show that much of this increased scatter is due to heating by starbursts which boost the far-infrared emission at 70\moo\ in a minority of objects (17$-$35\%). Overall this study supports previous work indicating AGN illumination as the dominant heating mechanism for MFIR emitting dust in the majority of low to intermediate redshift radio galaxies (0.03~$<z<$~0.7), with the advantage of strong statistical evidence. However, we find evidence that the low redshift broad-line objects ($z<$~0.1) are distinct in terms of their positions on the MFIR vs. [OIII] correlations. 

\end{abstract}


\section{Introduction}
\label{sec:intro}

Identifying the origin of prodigious thermal mid- to far-infrared (MFIR) emission is a key component for a comprehensive understanding of Active Galactic Nuclei (AGN). However, this task is not trivial, because the thermal MFIR emitting dust structures cannot be resolved in most of these galaxies. Therefore, past studies have favored a statistical approach to investigations, focusing on samples of radio galaxies, which can be selected without bias with respect to orientation (\citealp{golombek88}; \citealp{impey93} \citealp{heckman92,heckman94};  \citealp{hes95};  \citealp{haas04};  \citealp{shi05};  \citealp{cleary07}). Although these studies suggested that the mid-IR (5$-$30$\mu$m) emitting structures are heated by AGN illumination, the lack of sample completeness and the low mid-IR detection rate meant that the AGN heating hypothesis could not be supported with a full statistical analysis. 

Additionally, in the past, linking the active nucleus with the origin of the far-IR ($>$30\moo) emission from cool dust components proved difficult. The failure of uniform compact dust torus models to produce the observed far-IR SEDs of AGN  \citep{pier92} led to the proposal of  clumpy  dust torus geometries (\citealp{nenkova02,nenkova08}) which produce the required dust temperatures through cloud shadowing. Alternatively, other studies argued that the cool dust producing the far-infrared emission is predominantly heated by starbursts (\citealp{rowan95}, \citealp{schweitzer06}). However, the idea that starbursts dominate the heating of the far-IR emitting dust in AGN has yet to be firmly established with solid observational evidence. 

To address the problems associated with previous MFIR investigations of radio-loud AGN, that suffered from biased, incomplete and/or inhomogeneous samples, we carried out a program of deep Spitzer/MIPS MFIR photometric observations for a complete sample of 47 2Jy radio galaxies with redshifts 0.05~$<z<$~0.7 (Program 20233: PI Tadhunter). The results from these data are published in T07,  D08 and D09. The results have shown that [OIII] optical emission line luminosity ($L_{[\rm{OIII}]}$) is significantly correlated with both the mid- (24\moo) and far-infrared (70\moo) luminosities ($L_{24\mu m}$ and $L_{70\mu m}$ respectively). The AGN-photoionised narrow-line region (NLR) is emitted on a small scale ($\leq$5 kpc), therefore the $[\rm{OIII}]\lambda$5007 emission from the NLR is likely to provide a good indication of the intrinsic power of the illuminating AGN (e.g. \citealp{rawlings91}; \citealp{tadhunter98}; \citealp{simpson98} and discussion in D09). Consequently, the correlations between isotropic MFIR luminosity and [OIII] optical emission line luminosity provide strong empirical evidence to support AGN illumination as the dominant heating mechanism of the thermal MFIR emitting dust. Moreover, since radio-loud quasars, broad-line and narrow-line galaxies follow similar correlations between MFIR and [OIII] luminosities, without significant offsets between the two groups, the results also provide strong support for the orientation-based unified schemes for powerful, radio-loud AGN \citep{barthel89}.

In addition, we carefully considered the starburst contribution to the AGN heating of dust. We found that the objects showing optical evidence for starburst
activity from spectral synthesis modeling of their spectra, appear to have enhanced far-IR emission compared to the general sample. Our interpretation of these results is that, while AGN illumination is the primary heating mechanism for both the warm (mid-IR emitting, 24\moo) and cool (far-IR emitting, 70\moo) dust in most powerful radio galaxies, heating by starbursts acts to substantially boost the 70\moo\ luminosity in the 20$-$30$\%$ of objects in the 2Jy sample with optical evidence for star formation activity. 

The above results support the conclusions for previous studies of powerful radio galaxies (\citealp{heckman94};  \citealp{hes95};  \citealp{haas04};  \citealp{shi05};  \citealp{cleary07}) with the advantage of a thorough statistical analysis afforded to us by the complete and well detected sample. 
Having established these results for the 2Jy sample, which represents radio-loud AGN at intermediate redshifts (0.05~$<z<$~0.7), it is natural to investigate whether we find similar results for other samples of radio-loud AGN.

The low frequency ($\approx$170 MHz) selected 3C sample of radio-loud AGN has been favored by many previous investigators, because the low selection frequency means that it is unlikely to be affected by an orientation bias. Recently, deep optical spectroscopic data at both high and low resolution have been published for 3CR sources \citep{buttiglione09}, allowing us to create a sample that is complete in both MFIR and [OIII] observations. Furthermore, the 3CR objects make an ideal comparison to the higher frequency selected (2.7 GHz) 2Jy sample. Investigation of a low selection frequency sample allows us to test whether the selection frequency of the 2Jy sample leads to any biases that may affect our understanding of the MFIR emission from radio-loud AGN. 

We present here the analysis of Spitzer photometric observations for a complete sample of 3CRR radio galaxies \citep{laing83} with $z<$~0.1. The following investigation serves to test our previous conclusions concerning the origin of  the thermal MFIR, based on the southern 2Jy sample, using a sample of radio-loud objects with, on average, lower redshifts and radio powers, as well as a different selection frequency.

\section{Samples and data reduction}
\label{sec:sample}
 
\begin{deluxetable}{l@{\hspace{0mm}}c@{\hspace{0mm}}c@{\hspace{-2mm}}c@{\hspace{-2mm}}c@{\hspace{-2mm}}c@{\hspace{0mm}}c@{\hspace{0mm}}c@{\hspace{-2mm}}c@{\hspace{-2mm}}c}
\tabletypesize{\scriptsize}
\tablecaption{3CRR Sample Data \label{tbl-1}}
\tablewidth{0pt}
\tablehead{
\colhead{Name}{\hspace{0mm}} & \colhead{$z$}{\hspace{0mm}} & \colhead{RA(J2000)}{\hspace{-2mm}} &  \colhead{Dec(J2000)}{\hspace{-2mm}} &\colhead{Opt. Class}{\hspace{-2mm}} & \colhead{Rad. Class}{\hspace{-2mm}} & \colhead{SB}{\hspace{-2mm}} & \colhead{SB ref}{\hspace{-2mm}}
}
\startdata
3C33	&	\phantom{a}	0.060	&	\phantom{a}	01 08 52.8	&	 $+$13 20 14 	&	NLRG	&	FRII	&	No	&	6	\\
3C35	&	\phantom{a}	0.067	&	\phantom{a}	 01 12 02.2 	&	$+$49 28 35	&	WLRG	&	FRII	&	No	&	7	\\
3C98	&	\phantom{a}	0.030	&	\phantom{a}	03 58 54.4 	&	$+$10 26 03	&	NLRG	&	FRII	&	No	&	2	\\
DA240	&	\phantom{a}	0.036	&	\phantom{a}	07 48 36.9 	&	$+$55 48 58	&	WLRG	&	FRII	&	No	&	2	\\
3C192	&	\phantom{a}	0.060	&	\phantom{a}	08 05 35.0 	&	$+$24 09 50	&	NLRG	&	FRII	&	No	&	7	\\
4C73.08	&	\phantom{a}	0.058	&	\phantom{a}	09 49 45.9	&	 $+$73 14 23	&	NLRG	&	FRII	&	No	&	-	\\
3C236	&	\phantom{a}	0.101	&	\phantom{a}	10 06 01.7	&	 $+$34 54 10	&	WLRG	&	FRII	&	SB	&	1,4	\\
3C277.3	&	\phantom{a}	0.085	&	\phantom{a}	12 54 11.7 	&	$+$27 37 33	&	WLRG	&	FRI/FRII	&	No	&	8	\\
3C285	&	\phantom{a}	0.079	&	\phantom{a}	13 21 17.8 	&	$+$42 35 15	&	NLRG	&	FRII	&	SB	&	1,2	\\
3C293	&	\phantom{a}	0.045	&	\phantom{a}	13 52 17.8 	&	$+$31 26 46 	&	WLRG	&	FRI/FRII	&	SB	&	5	\\
3C305	&	\phantom{a}	0.042	&	\phantom{a}	14 49 21.6	&	 $+$63 16 14	&	NLRG	&	FRII/CSS	&	SB	&	5	\\
3C321	&	\phantom{a}	0.096	&	\phantom{a}	15 31 43.4 	&	$+$24 04 19	&	NLRG	&	FRII	&	SB	&	1,3	\\
3C326	&	\phantom{a}	0.090	&	\phantom{a}	15 52 09.1 	&	$+$20 05 24	&	NLRG	&	FRII	&	No	&	9	\\
3C382	&	\phantom{a}	0.058	&	\phantom{a}	18 35 03.4 	&	$+$32 41 47 	&	BLRG	&	FRII	&	U	&	10	\\
3C388	&	\phantom{a}	0.092	&	\phantom{a}	18 44 02.4	&	 $+$45 33 30 	&	WLRG	&	FRII	&	No	&	9	\\
3C390.3	&	\phantom{a}	0.056	&	\phantom{a}	18 42 09.0 	&	$+$79 46 17	&	BLRG	&	FRII	&	U	&	10	\\
3C403$^a$	&	\phantom{a}	0.059	&	\phantom{a}	19 52 15.7	&	 $+$02 30 23	&	NLRG	&	FRII	&	No	&	9	\\
3C445$^a$	&	\phantom{a}	0.057	&	\phantom{a}	22 23 49.6	&	$-$02 06 12	&	BLRG	&	FRII	&	U	&	10	\\
3C452	&	\phantom{a}	0.081	&	\phantom{a}	22 45 48.8 	&	$+$39 41 16	&	NLRG	&	FRII	&	no	&	7	\\

\enddata

\tablecomments{The basic parameters for the 3CRR sample are presented. Note
that 2 of the objects in the 3CRR sample are in common with the 2Jy
sample: 3C403 (PKS1949$+$02), 3C445 (PKS2221$-$02). Fluxes were
measured from Spitzer observations downloaded from the Spitzer
archive. Definitions for column 5 are: NLRG -- narrow-line radio galaxy, BLRG -- broad-line radio galaxy, WLRG -- weak-line radio galaxy.  Definitions for column 7 are: No -- No optical starburst, U -- Uncertain starburst objects, SB - Optical starburst objects. SB references are: (1) \citet{holt07}, (2) \citet{aretxaga01}, (3)
\citet{tadhunter96}, (4) \citet{odea01}, (5) \citet{tadhunter05}, (6)
\citet{robinson_thesis}, (7) \citet{wills02}, (8)
\citet{clark_thesis}, (9) Tadhunter, private communication, (10)
\citet{osterbrock76}.  Note that, as well as the evidence based on optical spectroscopy, the 
presence of energetically significant  star formation activity in 3C285, 
3C293, 3C321, and 3C305 is
supported by the detection of PAH features in their mid-IR Spitzer/IRS 
spectra (Dicken et al., 2010, in preparation; \citealp{shi07}). }

\end{deluxetable}

\begin{deluxetable}{l@{\hspace{1mm}}c@{\hspace{1mm}}c@{\hspace{1mm}}c@{\hspace{1mm}}c@{\hspace{1mm}}c@{\hspace{1mm}}c@{\hspace{1mm}}r@{\hspace{1mm}}c@{\hspace{1mm}}c}
\tabletypesize{\scriptsize}
\tablecaption{3CRR Sample Luminosities \label{tbl-2}}
\tablewidth{0pt}
\tablehead{
\colhead{Name}& {$z$}{\hspace{1mm}} & {$S_{24\mu m}(mJy)$}{\hspace{1mm}} &  {$\sigma$}{\hspace{-1mm}} &\colhead{$L_{24}$(W/Hz)}{\hspace{1mm}} & {$S_{70\mu m}(mJy)$}{\hspace{1mm}} &  {$\sigma$}{\hspace{-1mm}} & \colhead{$L_{70}$(W/Hz)}{\hspace{-3mm}} &\colhead{$L_{[\rm{OIII}]}$(W)}{\hspace{1mm}} & \colhead{$L_{radio}^{5GHz}$(W/Hz)} 
}
\startdata
 3C33	 	 	&\phantom{aa}	0.060	\phantom{a}&	 	 	99.4	&	 	0.2	&	 $	9.0\times10^{23	}$ 	&	 	\phantom{}	145.5		 	&	3.4	&	 $	1.3\times10^{24	}$ 	&	 $	\phantom{}	2.0\times10^{34	}$ 	&	 $	\phantom{}	3.4\times10^{25	}$	\\ 
 3C35	 	 	&\phantom{aa}	0.067	\phantom{a}&	 	 	0.9	&	 	0.2	&	 $	1.1\times10^{22	}$ 	&	 	\phantom{}	18.7		 	&	6.4	&	 $	2.3\times10^{23	}$ 	&	 $	\phantom{}	1.0\times10^{33	}$ 	&	 $	\phantom{}	6.5\times10^{24	}$	\\ 
 3C98	 	 	&\phantom{aa}	0.030	\phantom{a}&	 	 	45.5	&	 	0.6	&	 $	9.2\times10^{22	}$ 	&	 	\phantom{}	36.4		 	&	3.5	&	 $	7.4\times10^{22	}$ 	&	 $	\phantom{}	1.0\times10^{34	}$ 	&	 $	\phantom{}	7.0\times10^{24	}$	\\ 
 DA240	 	 	&\phantom{aa}	0.036	\phantom{a}&	 	 	3.9	&	 	0.4	&	 $	1.2\times10^{22	}$ 	&	 	\phantom{}	32.1		 	&	4.6	&	 $	9.7\times10^{22	}$ 	&	 $	\phantom{}	6.0\times10^{32	}$ 	&	 $	\phantom{}	5.2\times10^{24	}$	\\ 
 3C192	 	 	&\phantom{aa}	0.060	\phantom{a}&	 	 	6.3	&	 	0.4	&	 $	5.2\times10^{22	}$ 	&	 	\phantom{}	15.1		 	&	6.7	&	 $	1.3\times10^{23	}$ 	&	 $	\phantom{}	2.2\times10^{34	}$ 	&	 $	\phantom{}	1.6\times10^{25	}$	\\ 
 4C73.08	 	 	&\phantom{aa}	0.058	\phantom{a}&	 	 	44.6	&	 	0.4	&	 $	3.2\times10^{23	}$ 	&	 	\phantom{}	23.2		 	&	2.3	&	 $	1.7\times10^{23	}$ 	&	 $	\phantom{}	9.4\times10^{33	}$ 	&	 $	\phantom{}	4.5\times10^{24	}$	\\ 
 3C236	 	 	&\phantom{aa}	0.101	\phantom{a}&	 	 	17.3 &	 	0.3	&	 $	4.5\times10^{23	}$ 	&	 	\phantom{}	64.6	 	&	5.5	&	 $	1.7\times10^{24	}$ 	&	 $	\phantom{}	8.1\times10^{33	}$ 	&	 $	\phantom{}	4.1\times10^{25	}$	\\ 
 3C277.3	 	 	&\phantom{aa}	0.085	\phantom{a}&	 	 	9.0	&	 	0.3	&	 $	1.6\times10^{23	}$ 	&	 	\phantom{}	18.8		 	&	3.3	&	 $	3.4\times10^{23	}$ 	&	 $	\phantom{}	8.6\times10^{33	}$ 	&	 $	\phantom{}	2.1\times10^{25	}$	\\ 
 3C285	 	 	&\phantom{aa}	0.079	\phantom{a}&	 	 	46.2	&	 	0.4	&	 $	7.2\times10^{23	}$ 	&	 	\phantom{}	200.6 	&	2.7	&	 $	3.2\times10^{24	}$ 	&	 $	\phantom{}	3.6\times10^{33	}$ 	&	 $	\phantom{}	9.2\times10^{24	}$	\\ 
 3C293	 	 	&\phantom{aa}	0.045	\phantom{a}&	 	 	31.1	&	 	0.3	&	 $	1.5\times10^{23	}$ 	&	 	\phantom{}	303.0	 	&	6.7	&	 $	1.5\times10^{24	}$ 	&	 $	\phantom{}	6.4\times10^{32	}$ 	&	 $	\phantom{}	9.0\times10^{24	}$	\\ 
 3C305	 	 	&\phantom{aa}	0.042	\phantom{a}&	 	 	44.0	 &	 	0.1	&	 $	1.8\times10^{23	}$ 	&	 	\phantom{}	311.5		 	&	2.3	&	 $	1.3\times10^{24	}$ 	&	 $	\phantom{}	1.1\times10^{34	}$ 	&	 $	\phantom{}	4.3\times10^{24	}$	\\ 
 3C321	 	 	&\phantom{aa}	0.096	\phantom{a}&	 	 	264.0 &	 	0.1	&	 $	6.1\times10^{24	}$ 	&	 	\phantom{}	897.1	 	&	5.7	&	 $	2.1\times10^{25	}$ 	&	 $	\phantom{}	2.1\times10^{35	}$ 	&	 $	\phantom{}	2.7\times10^{25	}$	\\ 
 3C326	 	 	&\phantom{aa}	0.090	\phantom{a}&	 	 	0.7	&	 	0.1	&	 $	1.5\times10^{22	}$ 	&	 	\phantom{}	$<$9.0		 	&	-	&	 $	<2.0\times10^{23	}$ 	&	 $	\phantom{}	2.5\times10^{33	}$ 	&	 $	\phantom{}	9.5\times10^{24	}$	\\ 
 3C382	 	 	&\phantom{aa}	0.058	\phantom{a}&	 	 	98.8	&	 	0.2	&	 $	7.2\times10^{23	}$ 	&	 	\phantom{}	56.3	 	&	4.3	&	 $	4.1\times10^{23	}$ 	&	 $	\phantom{}	6.0\times10^{34	}$ 	&	 $	\phantom{}	1.8\times10^{25	}$	\\ 
 3C388	 	 	&\phantom{aa}	0.092	\phantom{a}&	 	 	2.6	&	 	0.2	&	 $	5.4\times10^{22	}$ 	&	 	\phantom{}	$<$11.1	 	&	-	&	 $	<2.5\times10^{23	}$ 	&	 $	\phantom{}	5.2\times10^{33	}$ 	&	 $	\phantom{}	3.8\times10^{25	}$	\\ 
 3C390.3	 	 	&\phantom{aa}	0.056	\phantom{a}&	 	 	217.1	&	 	0.2	&	 $	1.5\times10^{24	}$ 	&	 	\phantom{}	162.9		 	&	3.1	&	 $	1.1\times10^{24	}$ 	&	 $	\phantom{}	1.2\times10^{35	}$ 	&	 $	\phantom{}	3.2\times10^{25	}$	\\ 
 3C403$^a$	 	 	&\phantom{aa}	0.059	\phantom{a}&	 	 	193.0	&	 	0.2	&	 $	1.6\times10^{24	}$ 	&	 	\phantom{}	348.4		 	&	3.7	&	 $	2.7\times10^{24	}$ 	&	 $	\phantom{}7.2\times10^{34 }$ 	&	 $	\phantom{}	1.9\times10^{25	}$	\\ 
 3C445$^a$	 	 	&\phantom{aa}	0.057	\phantom{a}&	 	 	232.1	&	 	0.3	&	 $	1.7\times10^{24	}$ 	&	 	\phantom{}	186.4		 	&	5.2	&	 $	1.3\times10^{24	}$ 	&	 $	\phantom{}	1.7\times10^{35	}$ 	&	 $	\phantom{}	1.7\times10^{25	}$	\\ 
 3C452	 	 	&\phantom{aa}	0.081	\phantom{a}&	 	 	55.6	&	 	0.1	&	 $	4.3\times10^{23	}$ 	&	 	\phantom{}	55.7		 	&	4.7	&	 $	4.3\times10^{23	}$ 	&	 $	\phantom{}	1.1\times10^{34	}$ 	&	 $	\phantom{}	1.9\times10^{25	}$	\\ 
\enddata

\tablecomments{Columns 3,4,5,6,7 and 8 present the 24 and 70\moo\ fluxes, errors and luminosities
  for the 3CRR sample calculated from fluxes. Column 9 presents the [OIII] luminosities
  calculated from fluxes taken from \citet{buttiglione09},  except for the cases of 3C321, DA240 and 4C73.08 which were taken from \citet{saunders89}, and 3C445 taken from an average of data presented in \citet{osterbrock76}, \citet{tadhunter_thesis}, \citet{morris88} and  \citet{buttiglione09}. Column 10
  presents the 5 GHz radio luminosities taken from
  \citet{laing83}. Luminosities were again calculated using
  $H_{o}=71~km s^{-1} Mpc^{-1}, \Omega_{m}=0.27$ and
  $\Omega_{\lambda}=0.73$ along with spectral indices derived from the
  F(70)/F(24) flux ratios for the MFIR data, and the high frequency radio spectral index
  $\alpha_{2.7GHz}^{4.8GHz}$ for the radio data. }

\tablenotetext{a}{Also in the 2Jy sample}

\end{deluxetable}

This paper presents results for a complete sub-sample of 19 3CRR radio galaxies selected from the sample of  \citet{laing83} (see Table \ref{tbl-1}). We have limited these data for completeness to objects with FRII radio morphologies and redshifts $z\leq$~0.1. This leads to a sample with a high level of completeness in both Spitzer/MIPS detections and [OIII] $\lambda5007$ emission line flux measurements. In the following discussion we will refer to this sample as the 3CRR sample. Note that, although two objects in the sample (3C277.3, 3C293) have uncertain radio morphological classifications, and cannot  be confidently characterized as either FRI or FRII types, they are included here for completeness.

All of the 3CRR sample objects have been previously observed with Spitzer/MIPS. These data were downloaded as raw {\it MIPS} images from the Spitzer Reserve Observation Catalogue (ROC) and reduced in an identical wayto the 2Jy sample discussed in detail in D08. The MFIR flux and data and associated errors are presented in Table \ref{tbl-2} were extracted using aperture photometry, again using identical methods to those used for the 2Jy sample described in D08. The [OIII] fluxes were obtained from published deep optical spectra at both high and low resolution taken using Dolores on the TNG \citep{buttiglione09}, except DA240, 4C73.08, 3C321 and 3C445 a (see note in Table \ref{tbl-2}). We detect 100\% of the 3CRR sample at 24\moo\ and 89\% at 70\moo. 

For comparison we also discuss here the 2Jy sample from our previous study. This sample consists of 46 powerful radio galaxies and steep-spectrum quasars ($ F_{\nu} \propto \nu^{-\alpha},\alpha^{4.8}_{2.7} > 0.5 $) selected from the 2Jy sample of \citet{wall85} with redshifts 
0.05~$<z<$~0.7. A full discussion of the selection and MFIR data reduction for this sample is published in D08 along with tables of MFIR fluxes and luminosities (D09). Note that two objects overlap between the 3CRR and 2Jy samples (3C403, 3C445). 

In addition, published deep optical spectra have allowed us to identify the objects in the two samples with evidence for young stellar populations at optical wavelengths. The references for the stellar population analysis are given in Table \ref{tbl-1}.

\section{The Origin of the mid- to far-IR emission}
\label{sec:origin}

\begin{figure*}[p]
\epsscale{1}
\plotone{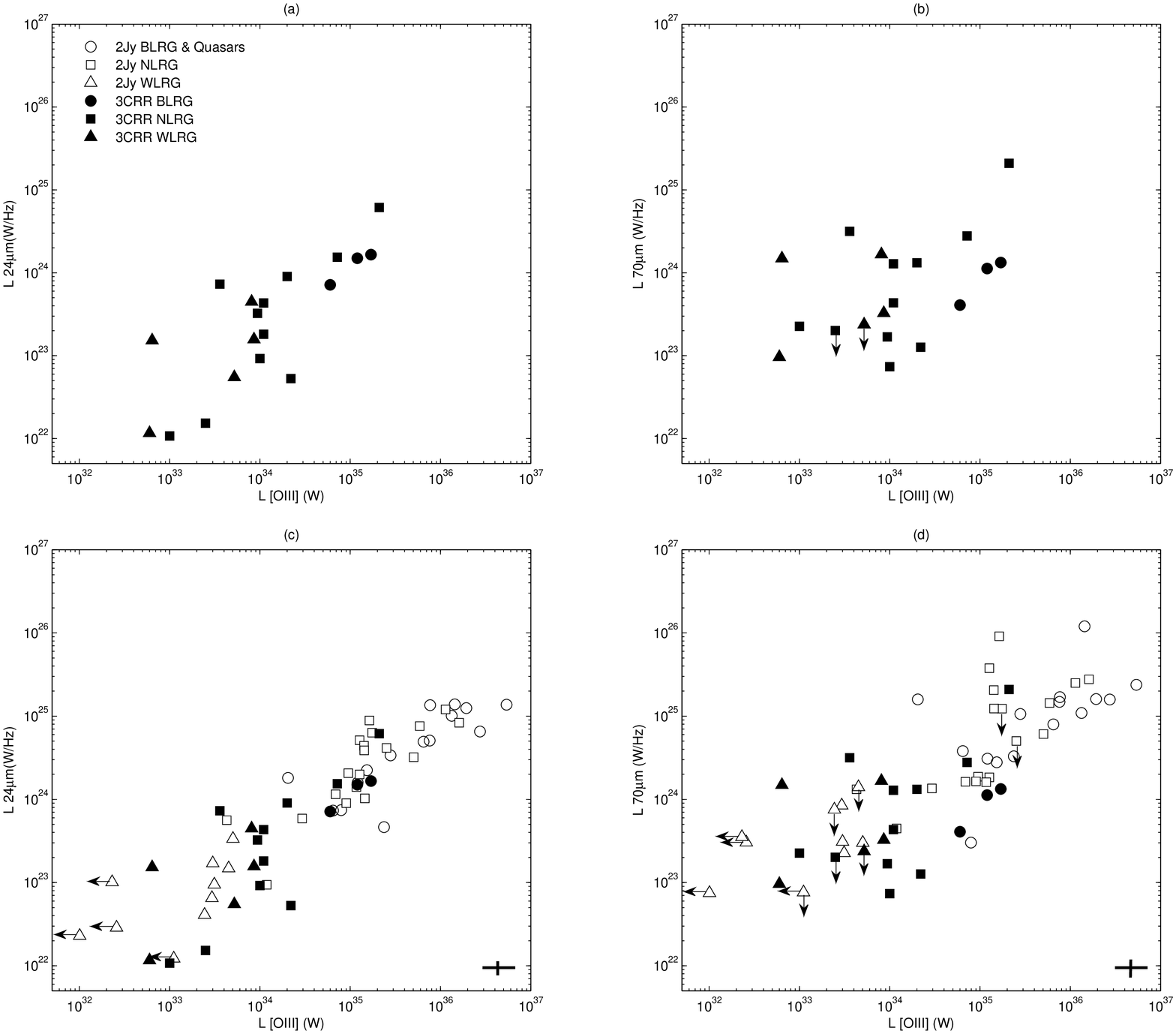}
\caption{: Luminosity correlation plots: (a) $L_{24\mu m}$
vs. $L_{[\rm{OIII}]\lambda 5007}$ and (b) $L_{70\mu m}$
vs. $L_{[\rm{OIII}]\lambda 5007}$ for the 3CRR sample alone. (c) $L_{24\mu m}$
vs. $L_{[\rm{OIII}]\lambda 5007}$ and (d) $L_{70\mu m}$
vs. $L_{[\rm{OIII}]\lambda 5007}$ for the combined 3CRR and 2Jy samples. \label{fig1} }
\end{figure*}

We first consider the 3CRR sample alone. In Figures \ref{fig1} (a) and (b) we plot the $L_{24\mu m }$ and $L_{70\mu m }$ against $L_{[\rm{OIII}]}$ for the 19 objects in the 3CRR sample. A visual inspection of Figure \ref{fig1} (a) identifies a correlation between $L_{24\mu m}$ vs. $L_{[\rm{OIII}]\lambda 5007}$. This correlation is statistically confirmed in Section \ref{sec:corr} and is consistent with the results found previously for the 2Jy sample (D09). Such a correlation supports the hypothesis that the warm, 24\moo-emitting dust is heated by direct AGN illumination, assuming that the [OIII] luminosity is a good indicator of intrinsic radiative AGN power. Secondly, a visual inspection of Figure \ref{fig1} (b), plotting $L_{70\mu m}$ vs. $L_{[\rm{OIII}]\lambda 5007}$, reveals less correlation in the 3CRR data compared to the result at $L_{24\mu m }$, also consistent with the results for the 2Jy sample (D09). 

We compare these results with the 2Jy sample in Figures \ref{fig1} (c) and (d) which show $L_{24\mu m }$ and $L_{70\mu m }$ plotted against $L_{[\rm{OIII}]}$  for the 3CRR and 2Jy samples plotted together.

Firstly, Figure \ref{fig1} (c) ($L_{[\rm{OIII}]}$ vs. $L_{24\mu m }$) reveals a strong correlation for the combined sample, with good continuity between the 3CRR and 2Jy samples at the low luminosity end of the correlation. Secondly, plotting $L_{70\mu m }$ vs. $L_{[\rm{OIII}]}$ for the combined sample also reveals a strong correlation, that is not apparent when plotting the 3CRR sample alone. Again these correlations are confirmed, statistically, in Section \ref{sec:corr}. 

 However, there is notable additional scatter in the $L_{70\mu m }$ vs. $L_{[\rm{OIII}]}$ correlation compared to that involving $L_{24\mu m }$. The crosses in the bottom right corners of Figures \ref{fig1} (c) and (d) show the maximum error for the points, demonstrating that the scatter is real and not purely a consequence of observational uncertainties (discussed further in section \ref{sec:far_origin}).  In this context, the apparent lack of a correlation between $L_{[\rm{OIII}]}$ and $L_{70\mu m }$ for the 3CRR sample alone is plausibly explained in terms of a combination of the high intrinsic scatter of the $L_{70\mu m }$ vs. $L_{[\rm{OIII}]}$ correlation and the small redshift and [OIII] luminosity range of the 3CRR sample. 

\section{Correlation statistics}
\label{sec:corr}

The high rate of Spitzer detections at MFIR wavelengths for the two samples allows us to conduct statistical tests on the significance of the correlations, discussed in the previous section and presented in Figure \ref{fig1}, using the Spearman rank correlation coefficient. However, although the overall detection rate is high for the observations of the two samples it is important to consider the effect of the 7 remaining upper limits in 70\moo\ luminosity and 4 upper limits in the [OIII] emission line luminosity. 

In order to remove the effect on the statistical tests of the 4 upper limits in [OIII], the 2Jy sample is limited to $z>$~0.06 and one object with an upper limit in [OIII] (PKS1839-48) was removed, leaving 38 of the original 46 2Jy objects. In addition, we applied a bootstrap method for dealing with six remaining 70\moo\ upper limits. For this we replaced the upper limits with 70\moo\ fluxes derived using the measured 24\moo\ flux of each object and a 70\moo/24\moo\ flux ratio chosen at random from the distribution of measured flux ratios for the detected sample objects. These 70\moo\ estimates were then converted to luminosities and included in the rank correlation test. This process was repeated 1000 times and the median of the correlation coefficients for those cycles was used for the correlation statistics involving 70\moo\ (see D09 for further details). Also, for the purposes of comparison, we investigated the correlations with upper limits using the ASURV (\citealp{isobe86}; \citealp{lavalley92}) package implemented in IRAF, including all upper limits in [OIII]. The survival analysis statistics used in ASURV  have been acknowledged as a powerful tool for analyzing samples with upper or lower limits. 

\begin{deluxetable}{l@{\hspace{0mm}}c@{\hspace{0mm}}c@{\hspace{0mm}}c@{\hspace{0mm}}c@{\hspace{0mm}}c@{\hspace{0mm}}c}
\tabletypesize{\scriptsize}
\tablecaption{3CRR and 2Jy Sample Statistical Analysis \label{tbl-3}}
\tablewidth{0pt}
\tablehead{ \colhead{Rank Correlation}{\hspace{0mm}} &
\colhead{$r_s$ (3CRR)}{\hspace{0mm}} &\colhead{significance}{\hspace{0mm}} &
\colhead{$r_s$ (3CRR$+$2Jy)}{\hspace{0mm}} &\colhead{significance}{\hspace{0mm}} &
\colhead{$r_s$ (ASURV)}{\hspace{0mm}} &\colhead{significance}{\hspace{0mm}}  }
\startdata 

(1) $L_{24}$ vs. $L_{[\rm{OIII}]}$       	&	0.74 &	$>99.5$\% &	0.89 &  $>99.9$\%	&	0.90 &  $>99.9$\%	\\
(2) $L_{70}$ vs. $L_{[\rm{OIII}]}$       	&	0.3 &	$>80$\% &	0.77 &  $>99.9$\%	&	0.76 &  $>99.9$\%	\\

\cutinhead{Partial rank correlation with \emph{z}}	
(3)$L_{24}$ vs. $L_{[\rm{OIII}]}$ & & & 0.78 & $>99.9$\% &  & \\
(4)$L_{70}$ vs. $L_{[\rm{OIII}]}$ & & & 0.47 & $>99.9$\% &  & \\

\enddata \tablecomments{Results of various
Spearman rank correlation statistics for the 3CRR plus the combined 3CRR
and 2Jy sample for the correlations presented in Figure \ref{fig1}. Values of $0<r_s<1$ are given for each test, where a
value close to 1 is highly significant. Columns 2 and 3 present the statistics for the 3CRR sample alone. Columns 4 and 5 present the statistics for the combined 2Jy and 3CRR sample undertaken
with a $z$ limited sample $z>$~0.06 for the 2Jy sample, but including all
the 3CRR sample; in this test the remaining upper limits were handled in a bootstrap method described in Section \ref{sec:corr}.  Columns 6 and 7 present statistics for all the objects in the combined sample, handling the upper limits using survival analysis statistics.}
\end{deluxetable}

The results of the Spearman rank tests for the correlations shown in Figure \ref{fig1} are presented in Table \ref{tbl-3}. As well as the percentage levels of significance, we also present the $r_s$ statistic, where a value of $r_s$ close to 1 is rated highly significant.

First, considering the 3CRR sample alone (Columns 2-3),
we find that the $L_{24\mu m}$ vs. $L_{[\rm{OIII}]}$ correlation is highly significant: we reject the null hypothesis that the variables are unrelated at a $>$99.5\% level. On the other hand, the $L_{70\mu m}$ vs. $L_{[\rm{OIII}]}$ correlation for the 3CRR sample alone is the least significant correlation we have examined: we only reject the null hypothesis that the variables are unrelated at a $>$80\% level. The additional scatter in the correlation between $L_{[\rm{OIII}]}$ vs. $L_{70\mu m}$ is further discussed in Sections \ref{sec:starburst} and \ref{disc}.

Inspecting the results for the combined 3CRR and 2Jy sample presented in Columns 4 and 6 of Table \ref{tbl-3} we find that all the tests show a correlation significance of better than 99.9\%, and the ASURV results reinforce those obtained using the bootstrap technique outlined above. These combined sample statistical tests strongly support the relation between the thermal MFIR emission and the [OIII] emission in radio galaxies for a broad range of redshift and radio powers. Because we believe that the [OIII] emission is a good indicator of AGN power, the combined 3CRR and 2Jy sample statistical results provide some of the strongest empirical evidence to date that the dominant heating mechanism for the MFIR continuum emission dust is AGN illumination.

Moreover, the second part of Table \ref{tbl-3} (rows 3 and 4) shows the results of a Spearman partial rank correlation test. This tests the hypothesis that the correlations are not intrinsic but arise because $L_{[\rm{OIII}]}$ and $L_{MFIR}$ are independently correlated with redshift. However in both cases the null hypothesis that the variables are unrelated is still rejected at the $>$99.5\% level of significance.

\section{Origin of the Far-Infrared Emission}
\label{sec:far_origin}

\begin{figure*}[p]
\epsscale{1}
\plotone{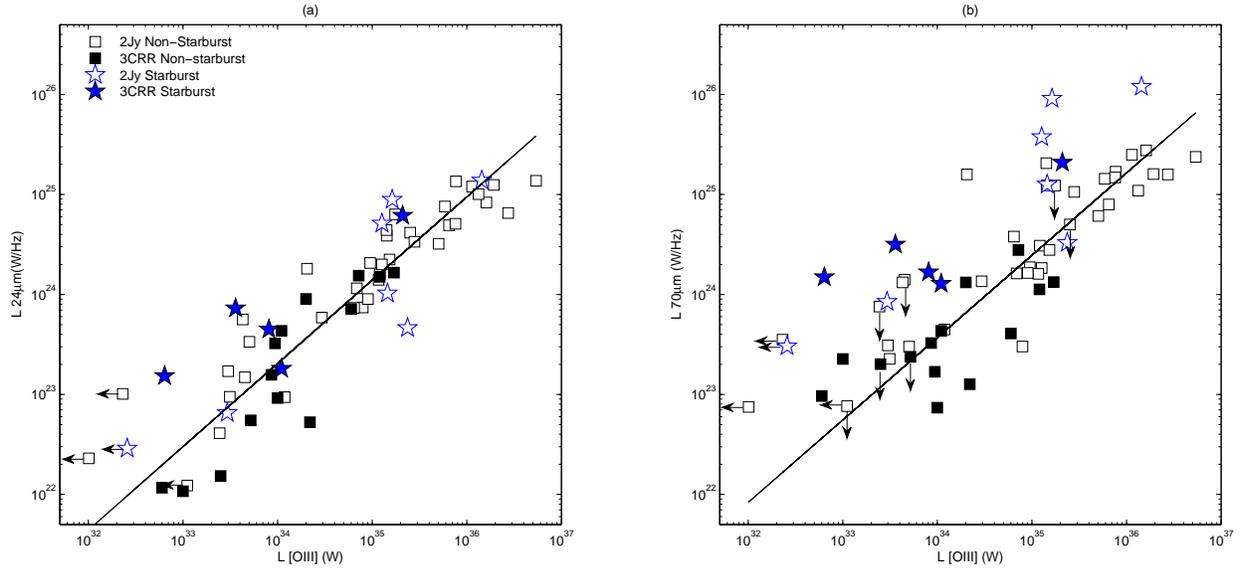}
\caption{: Plots showing the correlations
between MFIR and [OIII] luminosity for the combined 3CRR and 2Jy sample at (a) 24\moo\
and (b) 70\moo, with optical starbursts marked with separate
symbols (blue stars). The regression line is fitted to the entire 3CRR sample as well as the 2Jy sample objects with $z>$~0.06 in order to avoid most of the objects with
upper limits in [\rm{OIII}]. The fitting also does not include the 11
optical starburst objects and PKS1839$-$48 which has an upper limit in [OIII] flux.\label{fig2}}
\end{figure*}

We now consider the cause of the additional scatter in the correlations in the far-IR 70\moo\ luminosities. In particular, it is important to consider whether starbursts heat the cool dust  that radiates at far-infrared wavelengths, since morphological evidence suggests that at least some powerful
radio galaxies are triggered in major gas rich galaxy mergers
(e.g. \citealp{heckman86}). Such mergers are predicted to be associated with
powerful starbursts (e.g. \citealp{dimatteo05}). Moreover, understanding the
connection between starbursts and AGN is important for the
interpretation of sub-millimeter observations in the context of the star
formation history of radio-loud AGN at high redshift \citep{archibald01}. 

\subsection{Evidence for starburst heating in the far-IR continuum}
\label{sec:starburst}

By using results from our own spectral synthesis modeling work, as well as the literature, we have identified objects in both the samples that show clear evidence for recent star formation activity at optical wavelengths\footnote{In the following discussion we will
label these objects as \emph{optical starbursts}. However, we
emphasize that while some of these objects do contain genuine,
current, starburst activity, others are in fact in a
post-starburst phase. See \citet{tadhunter05,holt07,wills08} for further discussion.}  
(see Table \ref{tbl-1} for 3CRR and D09 for the 2Jy sample). Therefore, in Figure \ref{fig2} we  plot the $L_{24\mu m}$ and $L_{70\mu m}$ data against
$L_{\rm{[OIII]}}$ for the combined 3CRR and 2Jy sample, in this case
highlighting the 12 objects in the two samples that have been
identified as having optical starbursts. It is evident from a visual inspection of Figure \ref{fig2} that much of the additional scatter in the $L_{70\mu m}$ vs. $L_{[\rm{OIII}]}$ correlation compared to the $L_{24\mu m}$ vs. $L_{[\rm{OIII}] }$ correlation is a consequence of \emph{enhanced} far-IR emission in the optical starburst objects. This confirms the result for the 2Jy sample presented in T07 and D09, using an increased sample of starburst radio galaxies (a total of 12 optical starburst objects compared with the 7 in the 2Jy sample alone).

In order to evaluate the degree of enhancement in the far-IR emission above the main correlation, we have fitted regression lines on both plots (a) and (b) in Figure \ref{fig2}. The lines shown are the bisectors for objects without optical starburst of a linear least squares fits of $x$ on
$y$ and $y$ on $x$ (See Figure \ref{fig2} for details). On the 70\moo\ plot in Figure \ref{fig2} (b) it can be seen that 11 out of 12 of the optical starburst objects lie more than 0.3 dex (i.e. a factor of 2) above the regression line. 

We have also used a one dimensional Kolmogorov-Smirnoff (K-S) two sample test to compare the cumulative distributions of the vertical displacements from the fitted regression line in the $L_{70\mu m}$ vs. $L_{[\rm{OIII}]}$ plot (Figure \ref{fig2}). The test calculates the probability that the starburst and non-starburst objects are drawn from the same distribution. The null hypothesis that the optical starburst and non-optical starburst are draw from the same parent population is rejected at a better than 0.01\% level. This result further supports our interpretation that the far-infrared emission is boosted in the optical starburst objects.

\subsection{Color differences at low redshifts} 
\label{sec:blrg_under}

\begin{figure}[p]
\epsscale{0.5}
\plotone{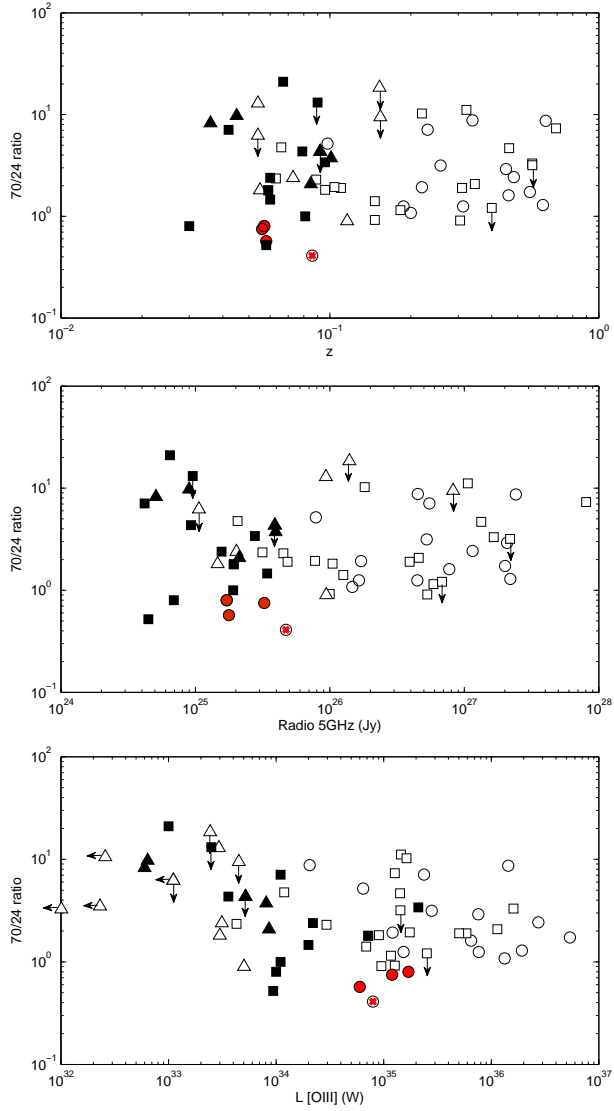}
\caption{:Plots of 70\moo/24\moo\ MFIR color vs. redshift, 5GHz total radio power and [OIII] luminosity (from top to bottom respectively). Symbols are the same as Figure 1. The four BLRG lying below the $L_{70\mu m} $ vs. $L_{[\rm{OIII}]}$  correlation with the warmest colors are marked with red. \label{fig3}}
\end{figure}

A detailed inspection of the  $L_{70\mu m} $ vs. $L_{[\rm{OIII}]}$ plots in Figures \ref{fig1} (d) and \ref{fig2} (b), identifies the six objects that lie on the bottom edge of the correlation (NLRG: 3C98, 3C192; BLRG:3C227, 3C382, 3C390.3, 3C445), the latter are displaced by 1-3 $\sigma$ based on the distribution of the residuals from the fitted regression line. The amount by which these objects lie below the regression line is much less than that by which the starburst objects are boosted above the correlation.  However, it is interesting that these objects are all at low redshift $z<$~0.09 and all (apart from 3C192) have much warmer colors than the rest of the objects in the 3CRR and 2Jy samples. Intriguingly, this group includes all four of the objects in the original papers that defined BLRG as a class (\citealp{osterbrock75} and \citealp{osterbrock76}) and notably all three of the broad line objects in the 3CRR sample. 

In Figure \ref{fig3} we present the 70\moo/24\moo\ color vs. redshift, $L_{5GHz}$ and  $L_{[\rm{OIII}]}$ for the 3CRR and 2Jy samples. In this figure we have marked in red the 4 BLRG that lie 1$-$2~$\sigma$ below the correlation between $L_{70\mu m} $ and $L_{[\rm{OIII}]}$. These 4 objects also have warm colors; $0.6< F(70)/F(24) < 0.8$ compared to a median of $F(70)/F(24)=2.3$. To begin with, Figure \ref{fig3} shows clearly that the 3CRR sample objects have, on average, lower redshifts and radio powers than the 2Jy sample. However, it is interesting that the 4 BLRG with warm colors all tend to higher $L_{[\rm{OIII}]}$ luminosity than all but two (3C321, 3C403) of the low redshift 3CRR objects. Indeed, the fact that these objects tend to higher [OIII] emission than objects of similar redshift and radio luminosity explains their position under the correlation between $L_{[\rm{OIII}]}$  and $L_{70\mu m} $. In addition, the fact that the  BLRG do not fall below the $L_{24\mu m}$ vs. $L_{[\rm{OIII}]}$  correlation is explained by an enhancement in their 24\moo\ emission as well as that in their [OIII] emission. This is consistent with their warm colors (see Fig 3), and the tendency of their Spitzer IRS spectra to peak at around 24\moo\ (Dicken et al. 2010 in prep).

\section{Discussion}
\label{disc}
\subsection{Origin of the mid- and far-infrared emission}
\label{sec:disc_origin}

From the plots of [OIII] emission line vs. MFIR luminosities, as well as from thorough statistical analysis, we have shown that the origin of the MFIR emission in 3CRR radio galaxies is most likely AGN illumination of the thermal emitting dust for the majority of objects. 

Considering the far-IR (cool dust) there are two main heating mechanism candidates: AGN illumination and starburst heating. However, the similar slopes of the 24 and 70\moo\ correlations (gradients of the fitted regression lines are 0.83 $\pm$ 0.05 and 0.82 $\pm$ 0.08 for 24\moo\ and 70\moo\ respectively\footnote{The slopes calculations do not include the identified optical starburst objects.}) presented in Figure \ref{fig2} indicate a common heating mechanism, for the warm and cool MFIR emitting dust components, i.e. AGN illumination. This is consistent with models that are capable of producing the broad MFIR SEDs by AGN illumination of near-nuclear clumpy tori (\citealp{nenkova02,nenkova08}). Alternatively, we showed in D09 that it is possible to account for the observed far-IR emission from the re-radiation of AGN illuminated narrow line clouds. Such a scenario is attractive as it does not require special or complex torus geometries for the circum-nuclear dust structures. 

Many previous studies have acknowledged the benefits of a statistical approach to understanding the origin of the thermal MFIR emission from radio-loud AGN, given the impossibility of resolving the MFIR emitting dust structures in the majority of objects. Such investigations began with exploratory IRAS studies investigating the contributions of thermal and non-thermal emission (\citealp{neugebauer86}; \citealp{golombek88}; \citealp{knapp90}; \citet{impey93}). However, it was the studies of \citet{heckman94} and \citet{hes95} that first provided evidence for a link between the MFIR emission and the AGN, finding 60\moo\ luminosities and extended radio luminosity to be correlated over 3 orders of magnitude. However, the detection rate in the far-IR of these IRAS observations was low ($<$30\%). Therefore \citet{heckman94} based their results on groups of objects averaged in redshift bins, and \citet{hes95} plotted only the objects that were detected. In further work \citet{haas98} suggested that the broad range of temperatures of dust emitting the MFIR emission argues in favor of AGN heating although this cannot rule out a starburst heating contribution. Subsequently, with ISO data, \citet{haas04} found the ratios of mid to far-infrared emission was higher for radio-loud AGN compared with ULIRGS. As AGN are likely to have hotter dust temperatures, this is consistent with, but does not prove, an AGN origin to the MFIR emission. 

One of the first studies to take advantage of sensitive MFIR data from Spitzer was that by \citet{shi05}. They found that a subset of the radio-loud AGN in their heterogeneous sample fall in the region of the MFIR color vs. [OIII]/H$\beta$ diagnostic diagram normally occupied by AGN (i.e. relatively warm colors and large [OIII/H$\beta$, see \citet{kewley01}), thus providing evidence that the cool dust is heated by AGN illumination rather than by starbursts. In addition, \citet{cleary07} found a correlation between MFIR luminosity (corrected for non-thermal contamination) and low frequency radio luminosity, suggesting AGN heating of the dust, based on a sample of 33 intermediate-redshift 3CR radio galaxies with a relatively low detection rate at far-IR wavelengths (60\%). The results we have presented are based on the combined 3CR and 2Jy complete sample of 63 objects with a 92\% detection rate at 70$\mu$m. Along with our previous work, these results strongly reinforce the idea that the heating of the cool, far-IR emitting dust in the majority of radio galaxies is dominated by AGN illumination.

\subsection{Starburst contribution to the far-infrared emission}
\label{sec:disc_starburst}

We have shown that the additional scatter above the main correlation between $L_{70\mu m} $ and $L_{[\rm{OIII}]}$  seen in 3CRR sample is accounted for by starburst boosting of the 70\moo\ far-IR emission. This enhancement is not seen for optical starburst objects at 24\moo. The results from the 3CRR sample add statistical weight to our previous study with only 7 optical starburst objects in the 2Jy sample to 12 optical starburst objects in the combined 3CRR and 2Jy sample. 

It is possible to estimate the rate of energetically significant
starburst activity in the 2Jy and 3CRR samples by considering the main optical
and infrared indicators of starbursts. The results are then: 12 (19\%) of the objects in the combined sample show unambiguous spectroscopic evidence for recent star formation activity at optical wavelengths; 12 (19\%) have cool MFIR
colors ($L_{70\mu m}/L_{24\mu m} > 5$); 22 (35\%) of the objects
lie more than 0.3 dex (factor $\times$2) about the regression line in
the $L_{70\mu m}$ vs. $L_{\rm{[OIII]}}$ correlation in Figure
\ref{fig2}; and 22 objects (35\%) show at least one of these indicators. Therefore an estimate of the proportion of powerful radio-loud AGN showing evidence for
energetically significant recent star formation activity in the combined
2Jy plus 3CRR sample is in the range 19$-$35\%. This result is also consistent with the study of \citet{shi07}, who find starburst tracing PAH features in only 2 of the 10 3CRR objects that overlap with the sample presented in this paper, and that of \citet{fu09} who do not find evidence for PAH features in any of the 12 objects in their FRII radio galaxies. 

It is generally accepted that the dust producing the continuum emission at 24\moo\ in AGN is heated almost exclusively by AGN illumination. In order for the cool dust emitting in the far-IR to be dominated by starburst rather than AGN heating, a remarkable degree of coordination between AGN
and starburst activity would be implied, given the strong correlations between $L_{24\mu m}$,
$L_{70\mu m}$ and $L_{[OIII]}$, and the similarity between the slopes of the
$L_{24\mu m}$ vs. $L_{[OIII]}$ and $L_{70\mu m}$ vs. $L_{[OIII]}$ correlations. 
Although such coordination cannot be entirely
ruled out, we consider it less likely. It is a fact that only a minority
of objects in the 2Jy and 3CRR samples show any evidence for recent star
formation activity, therefore, the current phase of AGN activity seen in these objects is unlikely to be  fueled by the gas flows that occur at the peaks of major gas-rich mergers. 

\subsection{Unified schemes}
\label{sec:disc_testing}

Thermal MFIR continuum emission can be used to test the orientation-based unified schemes for
powerful radio sources \citep[e.g.][]{barthel89}, under the assumption the MFIR emission is isotropic. For orientation-based unification to hold, correlations between MFIR emission and other isotropic wavelength emission such as low frequency radio or the [OIII], should reveal no differences between the relative positions of the different optical classes of objects. This tests the hypothesis that all the objects contain the same central engine, however, the optical classes arise because the optical emission is not emitted isotopically and can be obscured. In Figure \ref{fig1} we have labelled the objects by their optical class as broad-line radio galaxies and quasars (BLRG/Q), narrow-line radio galaxies (NLRG) and weak line radio galaxies (WLRG)\footnote{see \citet{tadhunter98} for definition and D09 for further detailed discussion.}. Applying such a test to the combined 3CRR and 2Jy sample in Figure \ref{fig1} (c) and (d), in general, little difference between the MFIR luminosities of BLRG/Q compared to NLRG is found. This result is in contrast to studies based on lower sensitivity IRAS data (\citealp{heckman94}, \citealp{hes95}), which suggested that BLRG/Q have enhanced MFIR emission  compared with NLRG objects of similar radio power.

However, as discussed above, it is apparent that {\it all} three 3CRR BLRGs and one of the two BLRG in the 2Jy sample with $z<$~0.1\footnote{Of the 5 low redshift ($z<$~0.1) BLRG in the combined sample, only PKS1733-56 ($z=$~0.098) falls above the correlation. However this object has a significant starburst component, as indicated by the detection of strong PAH features in its mid-IR spectrum (Dicken et al. 2010).} lie below the correlation between $L_{70\mu m}$ and $L_{\rm{[OIII]}}$. On the basis of Figure 3 it is likely that this displacement is due to their relatively strong [OIII] emission and not sub-luminous 70$\mu$m emission. The fact that this group of low-$z$ BLRG falls within the main body of the points in the $L_{24\mu m}$ vs. $L_{[OIII]}$ correlation is consistent with anisotropic [OIII] emission, provided that the 24$\mu$m emission is also enhanced in these objects by a similar degree to the [OIII]. Such enhancement is consistent with the warm colors of the low-$z$ BLRG, as well as the tendency of their Spitzer IRS spectra to peak at around 24\moo\ (Dicken et al. 2010 in prep). 

In order to reconcile the position of the low-$z$ BLRG in the MFIR correlation plots with the orientation-based unified schemes,  {\it both} the [OIII] emission and the 24$\mu$m emission must be anisotropic, and subject to significant dust extinction by the torus. Such anisotropy in mid-infrared emission above 15\moo\ has been seen for Seyfert galaxies when comparing the data from type 1 to type 2 \citep{buchanan06}. The required degree of anisotropy for the [OIII] emission is a factor of $\sim$3$-$7. In this case, a significant proportion of the [OIII] emission must be emitted on a relatively small ($\sim$pc) scale in these objects. For at least one of the low-$z$ BLRG $-$ 3C390.3 $-$ such a small scale for the NLR is supported by the evidence for significant variability in the narrow line CIV and [OIII] emission on a timescale of a few years (\citealp{clavel87}; \citealp{zheng95}). 

The fact that there is no difference between the relative positions of higher redshift ($>$0.1) BLRG/Q and NLRG in the of [OIII] emission line vs. MFIR luminosities plots supports the assumption that the [OIII] emission is isotropic for these objects in the combined sample. However, there is some evidence for {\it mild} anisotropy in the [OIII] emission of the -- generally higher redshift -- BLRG/Q in the 2Jy sample as discussed in detail in D09. However, the displacement for the 2Jy BLRG/Q below the $L_{70\mu m} $vs. $L_{[OIII]}$ correlation is not statistically significant (D09). Such a  difference could reflect a change in the properties of the torus and/or the spatial distribution of the NLR gas with the luminosity of the AGN.

\citet{vanbemmel01} compared ISO and IRAS photometric properties of 10 BLRG, with those of a heterogeneous sample of 5 NLRG detected by IRAS. They found that 7 of the BLRG objects have warm MFIR colors with continuum spectra peaking around 24$\mu$m, and explained this in terms of the BLRG objects lacking a cool dust components in the far-infrared --- an explanation supported by the apparent lack of morphological features due to dust in the BLRG relative to the NLRG in optical HST images\footnote{Note, however, that the NLRG sample of \citet{vanbemmel01} is heterogeneous and, even with the spatial resolution of the HST, it is difficult to detect near-nuclear dust features in the BLRG  because, unlike the NLRG, they have luminous point-like nuclei at optical wavelengths.}. If correct, this explanation would be inconsistent with the simplest versions of the orientation-based unified schemes, since it would imply that BLRG represent a separate class of radio-loud AGN that lack dust or, alternatively, a later evolutionary phase in the evolution of radio-loud AGN population. Based on the results presented in this paper (in particular, Figures 1 and 3), we feel that anisotropic [OIII] and 24$\mu$m emission is a more plausible explanation for  the differences between the properties of the low-$z$ BLRG and NLRG.

Unfortunately, the number of low-$z$ BLRG in our sample is small. To further investigate the apparently unusual MFIR properties of such objects, it will be important in the future to study observations of a larger sample, and also examine their mid-infrared spectra in more detail. 

\section{Conclusions}

In this paper we have investigated MFIR observations of a sample of 19 3CRR radio galaxies. The main conclusions are as follows.

\begin{itemize}

\item{From the statistical analysis of the 3CRR sample, correlating MFIR luminosities with the AGN power indicator [OIII], we conclude that the dominant heating mechanism for mid-IR emitting dust is AGN illumination. This result is consistent with our previous work based on the 2Jy sample of southern radio galaxies. Moreover, based on our analysis of the combined 2Jy and 3CRR sample, we conclude that the dominant heating mechanism for the cooler, far-IR emitting dust is also likely to be AGN illumination in the majority of radio-loud AGN. }

\item{Following the indications of previous work we have investigated whether additional scatter in the [OIII] vs. 70\moo\ luminosity correlation for 3CRR objects is a consequence of starburst heating which boosts the far-IR emission in some objects. We find that this is indeed the case for the 12 optically identified starburst objects in the combined 2Jy and 3CRR sample. We conclude that starburst heating of the far-IR emitting dust is important in only 17$-$35\% of objects. }

\item{Although we find no statistically significant differences between 
the properties  of the BLRG/Q and NLRG for the joint 2Jy and 3CRR sample, 
or  the 2Jy sample alone, we note that {\it all} the classical BLRG in our
3CRR sample at $z<$~0.1 show evidence for enhanced [OIII] emission and warmer 
MFIR colors compared with the majority of NLRG at similar redshifts. This suggests
that torus-induced anisotropy in [OIII] and 24\moo\
emission may be more significant in powerful radio galaxies at low 
redshifts than in their higher redshift counterparts. However, larger 
samples, along with a more detailed comparisons between the mid-IR spectra 
of BLRG and NLRG, are required to put this result on a firmer footing.}

\end{itemize}

\acknowledgments This work is based [in part] on observations made
with the Spitzer Space Telescope, which is operated by the Jet
Propulsion Laboratory, California Institute of Technology under a
contract with NASA. This research has made use of the NASA/IPAC
Extragalactic Database (NED) which is operated by the Jet Propulsion
Laboratory, California Institute of Technology, under contract with
the National Aeronautics and Space Administration. Based on
observations made with ESO Telescopes at the Paranal Observatory.
D. D. acknowledges support from NASA grant based on observations from Spitzer program 50588.

{\it Facilities:} \facility{Spitzer (MIPS)}.

\bibliographystyle{apj.bst} 
\bibliography{bib_list}

\end{document}